\documentclass[aps,pra,amsmath, twocolumn,showpacs,superscriptaddress]{revtex4-1}  

\usepackage{graphicx} 
\usepackage{subfigure}
\usepackage{float}
\floatstyle{boxed}
\usepackage{color}

\usepackage{ulem} %

\usepackage{amsmath}
\usepackage{amsfonts}
\usepackage{mathtools}\usepackage{braket}

\newcommand{\sgn}{\mathop{\mathrm{sgn}}\nolimits}
\newcommand{\tr}{\mathop{\mathrm{Tr}}\nolimits }
\newcommand{\N}{d}
\newcommand{\UD}{V}
\newcommand{\Up}{W}

\newcommand{\mfc}{{\mathfrak c}}
\newcommand{\mfs}{{\mathfrak s}}
\newcommand{\m}{m}

\def\ba{\begin{eqnarray}}
\def\ea{\end{eqnarray}}
\def\beq{\begin{equation}}
\def\eeq{\end{equation}}

\newcommand{\cc}[1]{{\color{blue}#1}}

\begin{document}

	\title{Spectrum estimation of density operators with alkaline-earth atoms}

	\author{Michael E. Beverland}
	\affiliation{Station Q, Quantum Architectures and Computation Group, Microsoft Research, Redmond, WA}
	
	\author{Jeongwan Haah}
	\affiliation{Station Q, Quantum Architectures and Computation Group, Microsoft Research, Redmond, WA}
	
	\author{Gorjan Alagic}
	\affiliation{Joint Center for Quantum Information and Computer Science, NIST/University of Maryland, College Park, MD 20742}
	
	\author{Gretchen K. Campbell}  
	\affiliation{Joint Quantum Institute, NIST/University of Maryland, College Park, MD 20742}
	
	\author{Ana Maria Rey} 
	\affiliation{JILA, NIST, and Department of Physics, University of Colorado Boulder, CO 80309}
	
	\author{Alexey V. Gorshkov}
	\affiliation{Joint Quantum Institute, NIST/University of Maryland, College Park, MD 20742}
	\affiliation{Joint Center for Quantum Information and Computer Science, NIST/University of Maryland, College Park, MD 20742}
	
	\date{\today}

	\begin{abstract}
		
		We show that Ramsey spectroscopy of fermionic alkaline-earth atoms in a square-well trap provides an efficient and accurate estimate for the eigenspectrum of a density matrix whose $n$ copies are stored in the nuclear spins of $n$ such atoms. 
		This spectrum estimation is enabled by the high symmetry of the interaction Hamiltonian, dictated, in turn, by the decoupling of the nuclear spin from the electrons and by the shape of the square-well trap.
		Practical performance of this procedure and its potential applications to quantum computing and time-keeping with alkaline-earth atoms are discussed.
		
	\end{abstract}
	
	\maketitle
	
	
	The eigenspectrum of a $\N$-dimensional density matrix $\hat \rho$ of a system characterizes the entanglement of the system with its environment \cite{horodecki09}.
	As it gives access to quantities such as purity, entanglement entropy, and more generally Renyi entropies, the eigenspectrum is an indispensable tool for studying many-body quantum states and processes in general and quantum information processors in particular \cite{eisert10b,nielsen00}.
	A strategy to estimate the spectrum specifies the measurements to be performed on $n$ copies of $\hat \rho$, along with a rule that specifies the estimated spectrum given measurement outcomes.
	It is natural that an optimal measurement should be invariant under arbitrary permutations [symmetry group $S_n$] and arbitrary simultaneous rotations [symmetry group $SU(\N)$] of all $n$ copies. 
	The well-known empirical Young diagram (EYD) algorithm involves a single joint entangled measurement on all $n$ copies which satisfies these symmetries, by projecting onto irreducible representations of $S_n \times SU(\N)$\cite{alicki88,SpectrumEstimationKeyl,Hayashi2002,Keyl2006,christandl06,odonnell15}.
	In this Letter, we show that Ramsey spectroscopy on $n$ fermionic alkaline-earth atoms stored together in a square trap can be used for spectrum estimation.
	We require each atom to have a copy of $\hat \rho$ stored in the $\N$-dimensional nuclear spin.
	Then spatially uniform Ramsey pulses between electronic states result in a joint measurement with $S_n \times SU(\N)$ symmetry, reminiscent of the EYD measurement. 
	
	Two unique features of fermionic alkaline-earth atoms are the metastability of the optically excited state $|e\rangle = {}^3\textrm{P}_0$ and the decoupling of the nuclear spin from the ($J=0$) electrons in both the ground state $|g\rangle = {}^1\textrm{S}_0$ and in $\ket{e}$.
	Thanks to these two features, alkaline-earth atoms have given rise to the world's best atomic clocks \cite{bloom14, nicholson15} and hold great promise for quantum information processing with nuclear and optical electronic qubits \cite{childress05,reichenbach07,hayes07,daley08,gorshkov09,daley11} and for quantum simulation of two-orbital, high-symmetry magnetism \cite{gorshkov10,cazalilla09,cazalilla14,zhang14,scazza14,cappellini14}.
	Spectrum estimation of $\hat{\rho}$, using a copy of $\hat{\rho}$ stored in the nuclear spin of each of $n$ $\ket{g}$ atoms, would be of great value in all of these applications.
	First, it can determine whether $\hat{\rho}$ describes a pure state, in which case the fermions would be identical and $s$-wave scattering  would not interfere with clock operation.
	Second, it can  be used to assess how faithfully the nucleus stores quantum information as one manipulates the electron \cite{childress05,reichenbach07,gorshkov09}.
	Finally, this procedure can be used to characterize the entanglement of a given nuclear spin with others in a many-atom state obtained via evolution under a  spin Hamiltonian \cite{honerkamp04,gorshkov10,cazalilla09,cazalilla14,zhang14,scazza14,cappellini14}; this would require $n$ copies of the many-atom state.
	
	\begin{figure}[b]
		\includegraphics[width=0.48\textwidth]{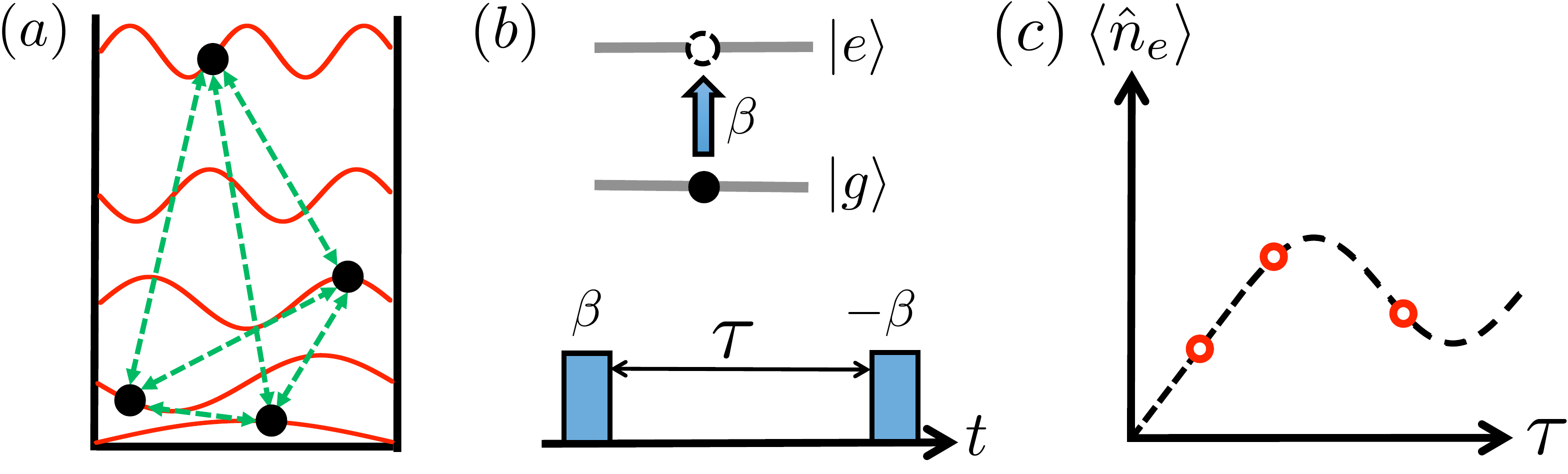}
		\caption{Spectrum estimation with alkaline-earth atoms.
			(a) $n$ copies of a $\N$-dimensional density matrix $\hat{\rho}$ are stored in the nuclear spin of $n$ fermionic alkaline-earth atoms
			trapped in a single square-well trap and prepared in their ground electronic state $\ket{g}$.
			(b) A Ramsey sequence is applied consisting of two pulses of area $\beta$ and $-\beta$, respectively,
			coupling $\ket{g}$ to the first excited electronic state $\ket{e}$.
			(c) The number $\langle \hat n_e \rangle$ of $e$ atoms is measured for different dark times $\tau$ (red circles)   
			between the pulses, allowing one to extract the spectrum of $\hat{\rho}$.
		}
		\label{fig:OverallProtocol}
	\end{figure}
	
	As illustrated in Fig.~\ref{fig:OverallProtocol}(a), to estimate the spectrum of $\hat \rho$, whose $n$ copies are stored in the nuclear spins of $n$ $\ket{g}$  atoms, we transfer all $n$ atoms into a single square well, with at most one atom per single-particle orbital.
	For sufficiently weak interactions, due to energy conservation and the anharmonicity of the trap, the $n$ occupied orbitals of the well remain unchanged throughout the experiment and play the role of individual sites. 
	Thanks to the decoupling of the $\N$-dimensional nuclear spin from the electrons, $s$-wave interactions give rise to a spin Hamiltonian with nuclear-spin-rotation $SU(\N)$ symmetry \cite{gorshkov10,cazalilla09}. Furthermore, the interaction strength between square-well orbitals labeled by positive integers $p \neq q$ is proportional to $\int_0^\pi \text{d} x \sin^2(p x) \sin^2(q x) = \pi/4$  and is thus independent of $p$ and $q$, giving rise to the site-permutation symmetry $S_n$  \cite{beverland14}. 
	Critically, the resulting Hamiltonian has $S_n \times SU(\N)$ symmetry.
	
	Remarkably, the independence of the interaction strength on $p$ and $q$ also makes the motional temperature of the atoms irrelevant.

	Our Ramsey protocol begins with the initial state of the $n$-atom system $|G \rangle \langle G| \otimes \rho^{\otimes n}$, where $\ket{G} = \ket{g\dots g}$ and each nuclear spin is in the same state $\hat{\rho}$.
	The first Ramsey pulse of area $\beta$ between $\ket{g}$ and $\ket{e}$ [Fig.\ \ref{fig:OverallProtocol}(b)] is implemented over short time $t_P = \beta/\Omega$ (so that interactions can be ignored), using Hamiltonian $\hat{H}_P = \tfrac{\Omega}{2} \sum_{k=1}^n \left( \hat{\sigma}_{eg}^k + \hat{\sigma}_{ge}^k\right)$ with Rabi frequency $\Omega$ and $\hat{\sigma}_{\mu \nu}^k = \ket{\mu}_k\!\!\bra{\nu}_{\! k}$. Since $s$-wave $e$-$e$ interactions are lossy \cite{zhang14}, we assume that the trapping of $\ket{e}$ atoms is temporarily loosened during the dark time $\tau$ \cite{daley08}, so that only $g$-$g$ interactions contribute via the spin Hamiltonian 
	\begin{eqnarray}
	\hat{H}_\textrm{D} = U \sum_{j < k} \hat \sigma_{gg}^{j} \hat \sigma_{gg}^{k}(1-\hat s_{jk}) - \delta \sum_k\hat{\sigma}_{ee}^k. \label{eq:Hamiltonian}
	\end{eqnarray}
	In the supplement we discuss the approach with a more general Hamiltonian \cite{supp}.
	Here $\hat s_{jk} = \sum_{r,r'=1}^d \ket{r}_j \!\! \ket{r'}_k \bra{r'}_j \!\! \bra{r}_k $ exchanges nuclear spins on sites $j$ and $k$ (so two identical fermions indeed do not $s$-wave interact), $\delta$ is the detuning of the Ramsey-pulse laser from the $g$-$e$ transition, $U=4 \pi \hbar a_{gg} \omega_\perp/L$, $a_{
		gg}$ is the $s$-wave $g$-$g$ scattering length, $L$ is the length of the square well, and $\omega_\perp$ is the frequency of the potential that freezes out transverse motion of the atoms \cite{beverland14}. After the second Ramsey pulse of area $-\beta$, the state is $\hat \rho' =  \hat \Up^\dagger \hat \UD \hat \Up | G \rangle \langle G | \hat  \rho^{\otimes n} (\hat \Up^\dagger \hat \UD \hat \Up)^\dagger$, where  $\hat \Up := \exp[-i t_P \hat H_P]$ and $\hat \UD := \exp[-i \tau \hat H_D]$.
	Finally, the number of $\ket{e}$ atoms  
	$\langle \hat n_e \rangle = \textrm{Tr}\left[ \hat n_e \hat \rho'\right]$ is measured, where $\hat{n}_e = \sum_j \hat{\sigma}_{ee}^j$. 
	
	We envisage starting with 
	$m \times R$ sets of $n$ atoms, each with nuclear spin state $\hat{\rho}$. 
	We denote the eigenspectrum of $\hat \rho$ as $\vec{p}=(p_1,p_2, \dots ,p_\N)$, ordered for future convenience as $p_1 \geq p_2 \geq \dots \geq p_\N$.
	For each dark time $\tau_1,\tau_2,\dots \tau_R$, we repeat the Ramsey protocol $m$ times and compute the average  [Fig.\ \ref{fig:OverallProtocol}(c)] to yield estimates of $\langle \hat n_e(\tau_1,\vec{p}) \rangle, \langle \hat n_e(\tau_2,\vec{p}) \rangle, \dots  \langle \hat n_e(\tau_R,\vec{p}) \rangle$. 
	Our key finding is that $\vec{p}$ can be inferred by fitting  the measured values to a pre-calculated expression of the mean number of $e$ atoms $\langle \hat n_e(\tau,\vec{p}) \rangle$.

	Although our approach is valid for all $n$, as $n$ increases, the distribution of measurement outcomes $\hat{n}_e/n$ becomes tightly peaked about its expectation value $\langle \hat n_e \rangle/n$  given by the  following expression in the large $n$ limit:
	\begin{eqnarray}
	\frac{\langle \hat n_e(\tau,\vec{p}) \rangle}{n}&=& \frac{\sin^2 \beta}{2}\left[ 1   - \sum_{r=1}^\N p_r \cos{(\omega_r \tau)} \right] + \tilde{\mathcal{O}} \!  \left(\frac{1}{\sqrt{n}}\right)\!\!,~~~~ \label{eq:NumberExcited3} 
	\end{eqnarray}
	where $\omega_r = U (n-1)  (1-p_r) \cos^2 \frac{\beta}{2} +\delta$.
	We use the notation that a tilde over the $\mathcal{O}$ indicates that we ignore logarithmic factors.
	Therefore the number of required repetitions $m$ decreases with $n$, making our approach particularly appealing in the regime of large $n$ [see Fig.~\ref{fig:ProbDist}(a)]. 
	
	The limiting cases of Eq.\ (\ref{eq:NumberExcited3}) are easily understood. Indeed, Rabi $\pi$-pulses ($\beta = \pi$) give zero since $\hat H_D \rightarrow -  n \delta$, so $\hat \Up^\dagger \hat \UD \hat \Up = \exp[i  n \delta \tau]$. Similarly, $\langle \hat n_e \rangle = 0$ in the absence of Rabi pulses ($\beta = 0$) since no $\ket{e}$ atoms are ever created. If $\hat \rho$ describes a pure state, in which case one of the $p_r$ is unity while the rest vanish, the interaction $U$ drops out (as it should for identical fermions) and we recover the familiar non-interacting expression. 
	
	\begin{figure}[b]
		\includegraphics[width=0.48\textwidth]{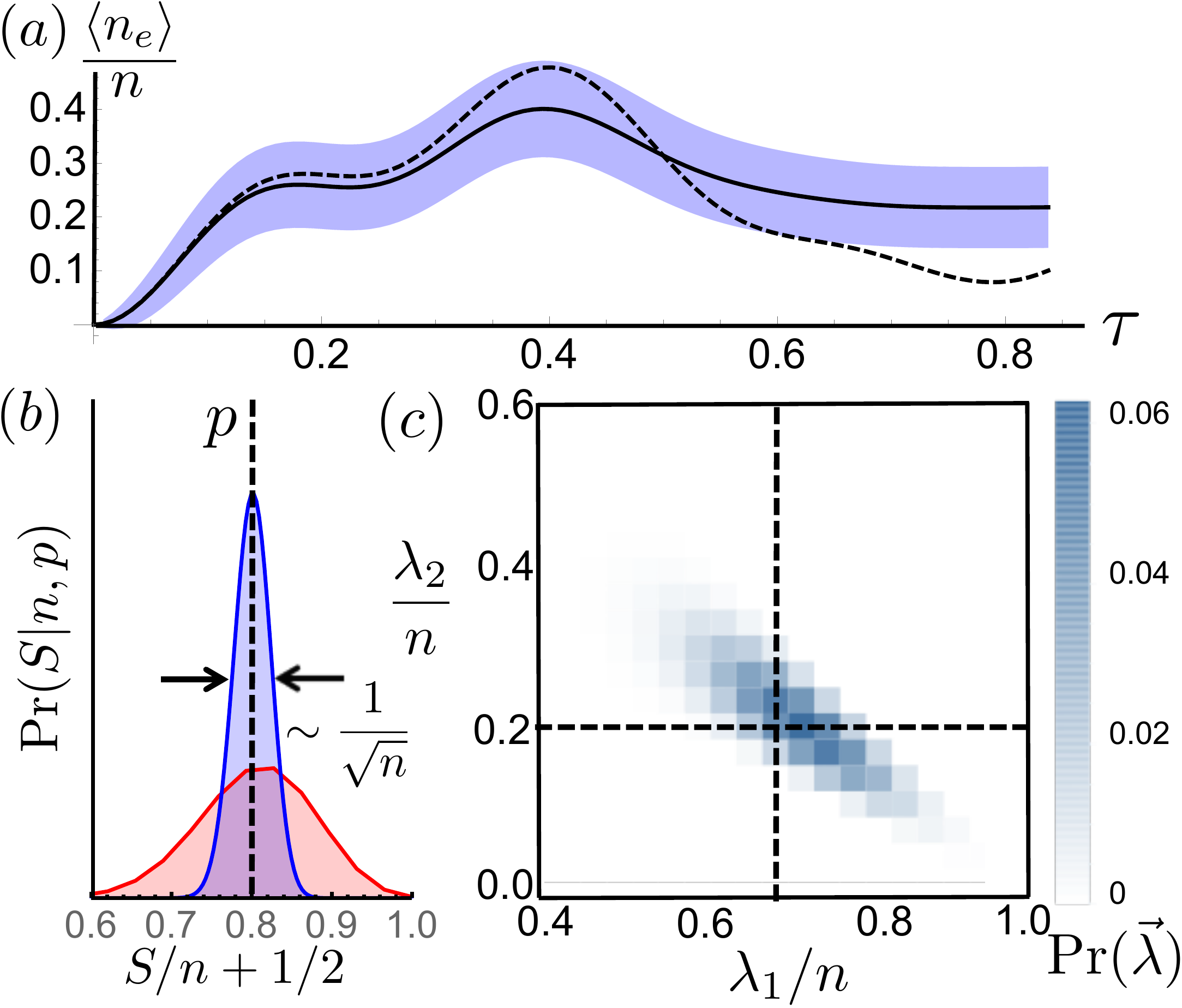}
		\caption{ 
			(a) For spectrum $\vec{p}=(0.7, 0.2, 0.1)$ and $n=30$, we compare the true expectation value $\langle \hat{n}_e(\tau,\vec{p}) \rangle/n$ (solid line) with that estimated using mean-field theory (dashed line). The blue region indicates outcomes that are within one standard deviation of $\langle \hat{n}_e(\tau,\vec{p}) \rangle/n$, where the standard deviation is estimated using the mean field result Eq.~(\ref{eq:meanfieldvariance}).  
			(b) The normalized probability distribution $\text{Pr}(S|n,p)$ for measurement outcome $S$ (and the estimate $S/n + 1/2$ for $p$) for $n=30$ and $n=300$ copies of  $\hat{\rho}$ with spectrum $(p,1-p)$ with $p = 0.8$.
			(c) For $n=30$, the probability distribution is shown for different outcomes $(\lambda_1, \lambda_2, n-\lambda_1-\lambda_2)$ given spectrum $(0.7, 0.2, 0.1)$.
		}
		\label{fig:ProbDist}
	\end{figure}
	
	
	\textit{EYD spectrum estimation.}---Before presenting the derivation of  the number of $e$ atoms, $\langle \hat n_e(\tau,\vec{p}) \rangle$, it is useful to review the original EYD spectrum estimation algorithm.
	For the familiar case of qubits ($\N=2$, or, equivalently, spin-$1/2$), the EYD algorithm can be stated as:
	
	Letting $(p,1-p)$ with $p \geq 1/2$ be the spectrum of $\hat \rho$, in the limit $n \rightarrow \infty$,
	a single measurement  on $\hat \rho^{\otimes n}$ of the total spin $\hat S^2$
	[with possible outcomes $S(S+1)$ with nonnegative $S = n/2, n/2 -1, \dots$]
	gives an outcome satisfying
	$p = 1/2 + S/n + \mathcal{O}(1/\sqrt{n})$. 
	
	This result follows from the fact that for large $n$ the measurement outcome distribution $\text{Pr}(S|n,p)$
	becomes peaked with mean and standard deviation $(p-1/2)n$ and $\sqrt{p(1-p)n}$ to leading order in $n$, as shown in Fig.~\ref{fig:ProbDist}(b) \cite{supp}.
	Note that the measurement operator $\hat S^2$ has symmetry group $S_n \times SU(2)$.
	The action of this symmetry group within each eigenspace of $\hat S^2$  corresponds one-to-one to a distinct irreducible representation of $S_n \times SU(2)$.
	
\begin{figure}[t]
		\includegraphics[width=0.48\textwidth]{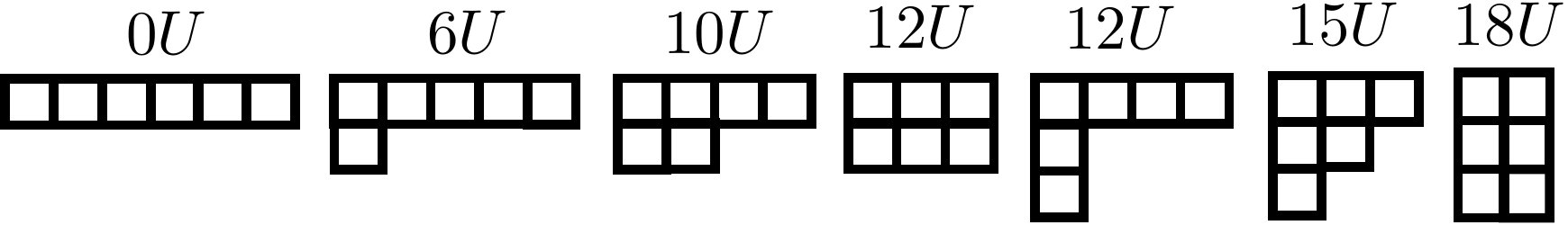}
		\caption{The Young diagrams $\vec{\lambda} = (\lambda_1, \lambda_2,...,\lambda_\N) $ for $n= 6$, $\N=3$.
			With all atoms in $\ket{g}$, the interaction Hamiltonian $\bra{G} \hat{H}_{D} \ket{G} = U\sum_{j < k} (1-\hat{s}_{jk})$
			has $S_n \times SU(\N)$ symmetry and is therefore diagonal in $\vec{\lambda}$-subspaces.
			The energy in $\bra{G} \hat{H}_{D} \ket{G}$ is displayed above each Young diagram. Notice two of the Young diagrams correspond to the same energy.
		}
		\label{fig:YoungDiagramExample}
	\end{figure}

	This generalizes to arbitrary $\N$. Thanks to Schur-Weyl duality~\cite{SchurTransform},
	the irreducible representations (irreps) of $S_n \times SU(\N)$ 
	in the $\N^n$-dimensional nuclear-spin Hilbert space $\mathcal{H}$ of $n$ atoms
	are in one-to-one correspondence with $\N$-row Young diagrams
	$\vec{\lambda} = (\lambda_1, \lambda_2,\ldots,\lambda_\N)$
	whose row lengths satisfy
	$\lambda_1 \geq \lambda_2 \geq \dots \geq \lambda_\N$ and $\sum \lambda_i =n$
	[see Fig.\ \ref{fig:YoungDiagramExample}].
	We write $\mathcal{H} = \bigoplus_{\vec{\lambda}} \mathcal{H}_{\vec{\lambda}}$,
	where the $\vec{\lambda}$-subspace $\mathcal{H}_{\vec{\lambda}} \subset \mathcal{H}$
	supports the $\vec{\lambda}$-irrep. 
	Any operator on $\mathcal{H}$ with $S_n \times SU(\N)$ symmetry
	has $\mathcal{H}_{\vec{\lambda}}$ as eigenspaces. 
	
	In the EYD algorithm, one measures the Young diagram on $\hat \rho^{\otimes n}$. 
	The distribution of outcomes $\text{Pr}(\vec{\lambda}|n,\vec{p})$  has a single peak near $n \vec{p}$  [see Fig.~\ref{fig:ProbDist}(c)]  with a typical deviation $\sum_{i} |  \frac{\lambda_i}{n} - p_i |$ of $\mathcal O( n^{-1/2})$ (for fixed $\N$)~\cite{christandl06}. 
	
	The experimental complexity associated with changing from the $S_n \times SU(\N)$ irrep basis $\mathcal{H}_{\vec{\lambda}}$ to the (generally easier to measure) computational basis makes implementing the EYD algorithm \cite{Bacon06} seem like a daunting task in practice. 
	The main result of this Letter is that the standard tool of Ramsey spectroscopy applied to fermionic alkaline-earth atoms in a square-well trap naturally accomplishes essentially the same task, allowing for efficient spectrum estimation.
	
	A hint at why our proposal achieves this goal is that the Hamiltonian restricted to the ground electronic state,
	$\bra{G} \hat{H}_{D} \ket{G}= U\sum_{j < k} (1-\hat{s}_{jk})$,
	is an operator on $\mathcal{H}$ with $S_n \times SU(\N)$ symmetry.
	Therefore $\bra{G} \hat{H}_{D} \ket{G}$ has subspaces $\mathcal{H}_{\vec{\lambda}}$ as energy eigenspaces, which can be probed by Ramsey spectroscopy.
	However the energies $E(\vec{\lambda}) = \frac{ U}{2} n(n-1)- \frac{U}{2} \sum_{i=1}^\N \lambda_i (\lambda_i -2i +1 ) $
	are not in one-to-one correspondence with subspaces $\mathcal{H}_{\vec{\lambda}}$ for $\N>2$  [see Fig.\ \ref{fig:YoungDiagramExample} for an example].
	Therefore, even if it were possible experimentally, direct measurement of the energy associated with 
	$\bra{G} \hat{H}_{D} \ket{G}$ 
	would not be sufficient to perform the EYD algorithm. 
	We will see that, remarkably, by accessing restrictions of $\hat H_D$ to different electronic states, Ramsey spectroscopy is powerful enough to uniquely identify $\vec{\lambda}$, thus enabling spectrum estimation.  
	
	
	\textit{Mean-field solution.}---To infer the spectrum, we need to calculate the Ramsey measurement expectation value,
	\begin{eqnarray} 
	\frac{\langle \hat n_e(\tau,\vec{p}) \rangle}{n} 
	&= & 
	\frac{\text{Tr} \left( \hat{\rho}^{\otimes n}~\hat{n}_e(\tau)   \right)}{n}, 
	\label{NumberExcitedNoVGeneralN2}
	\end{eqnarray}
	defining
	$\hat{n}_e(\tau) := \langle G | \Up^\dagger \UD^\dagger \Up  \hat{n}_e \Up^\dagger \UD \Up | G \rangle$, which is an operator on $\mathcal{H}$ with $S_n \times SU(\N)$ symmetry.
	We now show that, within the mean-field approximation, Eq.\ (\ref{NumberExcitedNoVGeneralN2}) can be evaluated using the expression in Eq.\ (\ref{eq:NumberExcited3}).
	
	Without loss of generality, we choose the eigenbasis of the initial nuclear-spin density matrix $\hat \rho$ as the nuclear spin basis. 
	At the mean-field level, time evolution under $\hat H_P$ and $\hat H_D$ does not create coherence between different nuclear spin states.  
	Let $\rho_{\mu \nu}^{r r}$ be the entry $\bra{\mu r} \hat \rho(\tau) \ket{\nu r}$ of the single-atom density-matrix $\hat \rho(\tau)$, where $\mu, \nu$ denote the electronic state ($g$ or $e$), while $r$ denotes nuclear spin. 
	Then the dark-time evolution keeps $\rho_{gg}^{r r}$ and $\rho_{ee}^{r r}$ unchanged, while
	\begin{eqnarray}
	\frac{\partial \rho_{eg}^{r r}}{\partial \tau}  = i \left[ \delta - U (n-1) \left( \rho_{gg}^{rr} - \sum_{r'} \rho_{gg}^{r' r'}\right)\right] \rho_{eg}^{r r}.
	\end{eqnarray}
	Putting this together with the two Ramsey pulses, we recover Eq.\ (\ref{eq:NumberExcited3}) without the $1/\sqrt{n}$ correction. 
	Since there is at most one $e$ atom in every site (spatial mode), the variance of $\hat n_e/n$ within the mean-field approximation is
	\begin{eqnarray}
	\langle (\hat n_e/n)^2  \rangle - \langle \hat n_e/n \rangle^2 = \frac{\langle \hat n_e/n \rangle -  \langle \hat n_e/n \rangle^2}{n}. \label{eq:meanfieldvariance}
	\end{eqnarray}
	This $1/\sqrt{n}$ standard deviation scaling is the same as that of the deviation of the mean-field value of $\langle \hat n_e(\tau,\vec{p}) \rangle/n$ from its exact value \cite{supp}. However the exact expression is still important for small $n$ which would occur when technical limitations prevent us from putting all available atoms into the same trap or when atoms are produced in small batches. In that case, we would need to repeat the experiment many times and will be sensitive to the deviation of the meanfield value from the exact result. Therefore, we now evaluate Eq.\ (\ref{NumberExcitedNoVGeneralN2}) exactly.
	
	
	\textit{Exact solution}.---To avoid clutter, we drop hats on operators and arrows on vectors and introduce abbreviations:
	$\mathfrak c := \cos \frac{\beta}{2}$, $\mathfrak s := \sin \frac{\beta}{2}$.
	We define a basis $\ket{E}$ of binary vectors,
	$E = (E_1,E_2,...,E_n) \in \{0,1\}^n$, 
	where the $k$th atom is in electronic state $\ket{g}$ ($\ket{e}$) when $E_k=0$ ($E_k=1$). 
	We also denote by $|E|$ the number of $1$'s in $E$. Expanding $\Up \ket{G}$ in the $\ket{E}$ basis,
	\begin{align}
	n_e(\tau)
	&= \sum_{\mathclap{E',E \in \{0,1\}^n}} i^{|E'|-|E|}\mfc ^{2n-|E|-|E'|} \mfs^{|E|+|E'|} \nonumber \\
	&\quad \quad \times \bra{E'} \UD ^\dagger \Up  n_e \Up^\dagger \UD \ket{E}.
	\label{SumForNetau}
	\end{align}
	Since $\Up  n_e \Up^\dagger$ is a sum of single-atom operators, 
	terms in which strings $E$ and $E'$ differ on more than one site vanish.
	When $E'=E$,
	\begin{align}
	\bra{E} \UD^\dagger \Up  n_e \Up^\dagger \UD \ket E 
	&= 
	\bra{E} \Up  n_e \Up^\dagger \ket{E} \nonumber \\
	&= (n-|E|) \mfs^2 + |E| \mfc^2,
	\end{align}
	since $\UD \ket{E} = e^{i\delta |E| \tau} \exp \left[{-i\alpha \sum_{j < k \notin E} (1-\hat s_{jk})}\right] \ket{E}$.
	Here $\alpha = U \tau$, $j < k \notin E$ is a sum over all pairs $j < k$ such that $E_j=0$ and $E_k=0$.
	Terms with $E'=E$ thereby sum to $2n \mfc^2 \mfs^2  = \frac{n}{2} \sin^2 \! \beta$ in Eq.~\eqref{SumForNetau}.
	
	When $E'$ and $E$ only differ on the $k$th atom such that $E_k=1$ and $E'_k=0$,
	\begin{eqnarray}
	\!\!\!\!
	\bra{E'} \! \UD^\dagger \Up  n_e \Up^\dagger \UD \! \ket E \! &=& -i \mfc \mfs 
	e^{i \delta \tau} \! \underbrace{ e^{i \alpha  \sum_{j \notin E } (1-s_{jk}) } }_{\mathcal A _E} , 
	\end{eqnarray}
	as 
	$e^{-i \alpha \sum_{j < l \notin E' } s_{jl} } e^{i \alpha \sum_{j < l \notin E } s_{jl} }=e^{-i \alpha \sum_{j \notin E } s_{jk} }$,
	which holds since the exponents commute.
	Defining $\mathcal A _E$ as given by the underbrace, the contribution to the sum in Eq.~\eqref{SumForNetau} of $E$ and $E'$ that differ on a single atom is
	\begin{align}
	\!\!\!\!\! -\sum_{k=1}^n \!\!\!\!\!\!\!\! \sum_{~~~~\tiny \begin{matrix}E \! \in \! \{0,1\}^{n} \\ E_k \!=\! 1 \end{matrix}} \!\!\!\!\!\!\!\!\!\! \mfc^{2n-2|E|+2} \mfs^{2|E|}
	\text{Tr} \! \left[  \rho^{\otimes n}  ( e^{i \delta \tau} \!\! \mathcal{A}_E  +  e^{-i \delta \tau} \!\! \mathcal{A}^\dagger_E )  \right] \!.
	\end{align}
	Note that $\text{Tr}( \rho^{\otimes n} \mathcal{A}_E )$ is invariant under site permutation, and therefore depends only on $|E|$.
	For integer $w = 0,1,...,n-1$, define the convenient $|E|=w+1$ representative operator $\mathcal{B}_w := e^{i \alpha \sum_{j =1}^{n-w-1}(1-s_{jn}) } $.
	Then,
	\begin{eqnarray}
	\!\!\!\!\!\!\! \frac{\langle \hat n_e(\tau,\vec{p}) \rangle}{n} ~  && = \nonumber \\
	&& \!\!\!\!\!\!\!\!\!\!\!\!\!\!\!\! \frac{ \sin^2 \!\beta}{2} 
	\left[ \! 1 \! -  \!  \sum_{w=0}^{n-1} \Pr(w | n,\beta)~  \mathfrak{Re}\, \{ e^{i \delta \tau}   \text{Tr}( \rho^{\otimes n} \mathcal{B}_w  ) \} \! \right] \!\! , ~~
	\label{neforarbitraryn}
	\end{eqnarray}
	where 
	$\Pr(w|n,\beta) := \binom{n-1}{w}  \mfc^{2(n-w-1)} \mfs^{2w}$
	is the binomial distribution obtained from expanding 
	$(\mfs^2 + \mfc^2)^{n-1}=1$.
	
	We evaluate $ \text{Tr}( \rho^{\otimes n} \mathcal{B}_w  )$ in two ways. The first way (presented below) uses group representation theory and illustrates the connection to the EYD algorithm, and yields an expression that can be evaluated conveniently numerically. The second approach (provided in the supplement \cite{supp}), is used to prove that the asymptotic result in Eq.~(\ref{eq:NumberExcited3}) deviates from the exact result by $\tilde{\mathcal{O}}(1/\sqrt{n})$.
	
	As $\text{Tr}( \rho^{\otimes n} \mathcal{B}_w  )$ is invariant under $S_n\times SU(\N)$ actions,
	\begin{eqnarray}
	\text{Tr}( \rho^{\otimes n} \mathcal{B}_w)    = \sum_\lambda \text{Pr}(\vec{\lambda}|n,\vec{p}) \text{Tr}_\lambda( \mathcal{B}_w)  ,
	\label{eq:BandEYD}
	\end{eqnarray}
	where $ \text{Pr}(\vec{\lambda}|n,\vec{p})$ is the EYD probability distribution, and $\text{Tr}_\lambda$ is a trace over the $\lambda$-subspace $\mathcal{H}_{\lambda}$.
	Now we show 
	\begin{eqnarray}
	\!\!\! \text{Tr}_\lambda( \mathcal{B}_w) \! = \!e^{i\alpha(\!n\!-\!w\!-\!1\!)} \!\!
	\sum_\xi \! \Pr(\xi|w,\lambda) \! \sum_{r=1}^\N \!\frac{\| \xi^{-r} \| }{\| \xi \|}
	e^{-i\alpha(\!\xi_r \! - r \!)} \! , ~~~
	\label{Aexpression}
	\end{eqnarray}
	where the sum is over all irreps $\xi$ of $S_{n-w}$, and $\text{Pr}(\xi|w,\lambda) := \frac{m(\lambda,\xi) \|\xi \|}{ \| \lambda \|}$ is a probability distribution defined in terms of the multiplicity $m(\lambda,\xi)$ of irrep $\xi$ of $S_{n-w}$ when regarding $\lambda$ as a (reducible) representation of the subgroup $S_{n-w} \subset S_{n}$. 
	For an irrep $\mu$ of $S_m$, its dimension is denoted $\| \mu \|$, the length of the $r$th row is $\mu_r$, and $\mu^{-r}$ is an irrep of $S_{m-1}$ defined by removing a box from the $r$-th row of $\mu$.
	
	To begin, note $\mathcal{B}_w$ is composed of permutations 
	in the subgroup $S_{n-w}$ of the first $n-w-1$ sites, along with the $n$th site.
	From this observation, we regard the representation space $\lambda$ as a representation of $S_{n-w}$,
	to obtain a \textit{reducible} representation $\lambda |^n_{n-w}$ of $S_{n-w}$.
	Note that we ignored the $SU(\N)$ Hilbert space and considered $S_n$ alone since $\mathcal{A}_E$ is written in terms of elements of $S_n$, which are each themselves $SU(\N)$ symmetric.
	This decomposes into a direct sum of irreps $\xi$ of $S_{n-w}$ 
	as $\lambda|^n_{n-w} \cong \bigoplus_\xi m(\lambda, \xi) \xi$.
	The multiplicity $m(\lambda,\xi)$ is the number of distinct paths from $\lambda$ to $\xi$, where each step in a path is a Young diagram, with one box removed from the previous step \cite{supp}.
	Since $\mathcal{B}_w$ is invariant under permutation of the first $n-w-1$ sites,
	we can finally diagonalize $\mathcal B_w$ by further restricting
	each $\xi$-irrep of $S_{n-w}$ to subgroup $S_{n-w-1} \subset S_{n-w}$;
	$\mathcal{B}_w$ must have each $\xi^{-r}$-subspace as an eigenspace.
	The eigenvalue of the $\xi^{-r}$-subspace is $ e^{i\alpha(\!n\!-\!w\!-\!1\!)} e^{-i \alpha (\xi_r-r)}$~\cite{supp},
	resulting in Eq.~\eqref{Aexpression}.
	
	We have introduced three probability distributions
	$\Pr (\lambda|n,p)$, $\Pr (w|n,\beta)$, and $\Pr (\xi|w,\lambda)$,
	all of which turn out to be unimodal for large $n$.
	In the large $n$ limit, the unimodality together with the fact that $\frac{\|\xi^{-r}\|}{\|\xi\|} \to \frac{\xi_r}{\sum_j \xi_j}$ recovers the mean field result Eq.\ (\ref{eq:NumberExcited3}).
	For $n$ and $d$ which are too large to evaluate $\langle \hat n_e(\tau,\vec{p}) \rangle /n$ exactly, one can still obtain a more precise estimate with this approach than that given by Eq.\ (\ref{eq:NumberExcited3}) by dropping terms associated with negligible contributions to the distributions \cite{supp}.
	
	
	\textit{Experimental considerations.}---In Ref.~\cite{beverland14} we suggest an implementation to trap tens of $^{87}$Sr atoms in a square well potential
	by freezing out the $x$ and $y$ directions using a strong red-detuned laser such that $\omega_\perp = 2 \pi \times 10$ kHz, and ``capping'' the ends of the tube of length $L \sim 10 \mu$m with a blue-detuned laser.  These parameters and the s-wave $^{87}$Sr scattering length $a_{gg} = 5.1$ nm \cite{Escobar08} result in $U =(4 \pi a_{gg} \omega_\perp)/L \approx 2 \pi \times 10$ Hz, allowing one to trap $\lesssim 20$ atoms. 
	
	The relevant timescale for Eq.(2) is $1/(nU) \sim 1$ ms.
	
	One can use a  build-up cavity to increase barrier height of the caps and $\omega_\perp$, allowing one to trap  more atoms and therefore carry out higher-resolution  spectrum estimation.
	
	To avoid losses caused by $e$-$e$ collisions, we propose temporarily loosening the $e$ trap during the dark time, which is readily doable for our choice of internal states \cite{daley08}. This should be performed slowly with respect to $\omega_\perp$ and quickly with respect to $U$.
	
	An experimentally simpler approach is to use $\beta$ sufficiently small as to make $e$-$e$ interactions negligible; this will, however, decrease the signal requiring additional repetitions of the experiment. 
	In the supplement, we include $e$-$g$ collisions in the mean-field treatment \cite{supp}. We include analysis of experimental imperfections in the supplemental material \cite{supp}.
	
	\textit{Outlook.}---We have shown that alkaline-earth atoms can be used as a special-purpose quantum computer capable of measuring the spectrum of a density matrix, motivated by EYD. 
	It is possible that many other useful quantum information tasks can be accessed in similar systems with special symmetry properties.
	In particular, an important extension of our work would be to find an efficient implementation of full-state tomography in current experimental systems. On the other hand, it would also be interesting to know if one can improve on our proposal if one seeks to measure a simpler quantity than the full spectrum \cite{odonnell15},
	such as the purity. 
	
	\begin{acknowledgments}
		\textit{Note.}---While finalizing the manuscript, we learned of a proposal \cite{pichler16} to perform spectrum estimation with Rydberg atoms  
		using a sequence of swap operations between two copies of the system, controlled by an ancilla.
		
		We thank S. P.\ Jordan, J.\ Preskill, K. R. A.\ Hazzard,  M.\ Foss-Feig, P.\ Richerme, M.\ Maghrebi, and R.\ de Wolf for discussions.
		This work was supported by 
		NSF JQI-PFC,
		NSF IQIM-PFC-1125565,
		NSF JILA-PFC-1734006,
		NSF-QIS, NIST, ARO, ARO MURI, ARL CDQI, DARPA (W911NF-16-1-0576 through the ARO), AFOSR, 
		and the Gordon and Betty Moore foundations.
		MEB and AVG acknowledge the Centro de Ciencias de Benasque Pedro Pascual for hospitality.
		JH is supported by Pappalardo Fellowship in Physics at MIT.
	\end{acknowledgments}

\begin{widetext}
	
	\renewcommand{\thesection}{S\arabic{section}} 
	\renewcommand{\theequation}{S\arabic{equation}}
	\renewcommand{\thefigure}{S\arabic{figure}}
	\setcounter{equation}{0}
	\setcounter{figure}{0}
	
	\section{Mean-field theory}
	\def\e{\epsilon}
	\def\d{\downarrow}
	\def\u{\uparrow}
	
	\def\e{\mathcal{E}}
	
	Here we derive Eq.\ (2) in the main text, when elastic $e-g$ and $e-e$ interactions are included in the dark Hamiltonian $H_D$. 
	We remind the reader that the mean-field analysis is not valid for small $n$.
	Without loss of generality, let's work in the nuclear spin basis, where the initial nuclear-spin density matrix is diagonal: $\rho^{m m'} = \delta_{m m'} p_m$.  After the first pulse, we then have
	\ba
	\rho_{gg}^{m m'} &=& \delta_{m m'} p_m \cos^2(\beta/2),\\
	\rho_{ee}^{m m'} &=& \delta_{m m'}  p_m \sin^2(\beta/2),\\
	\rho_{ge}^{m m'} &=& \delta_{m m'}  p_m \frac{i}{2} \sin \beta.
	\ea
	The generalized dark Hamiltonian is \cite{gorshkov10}
	\ba
	\hat{H}_D &=& U_{gg} \sum_{j < k} \hat{\sigma}^j_{gg} \hat{\sigma}^k_{gg} (1 - s_{jk})  +U_{ee} \sum_{j < k} \hat{\sigma}^j_{ee} \hat{\sigma}^k_{ee} (1 - s_{jk}) - \delta \sum_k \hat{\sigma}^k_{ee} \label{longeq} \nonumber \\
	&& + V n_{g} n_{e} +V \sum_{p q, j \neq k}  \hat{c}^\dagger_{j e p} \hat{c}_{k e p} \hat{c}^\dagger_{k g q} \hat{c}_{j g q} + V^{ex} \sum_{p q, j  k} \hat{c}^\dagger_{j g p} \hat{c}^\dagger_{k e q} \hat{c}_{j g q} \hat{c}_{k e p} + V^{ex} \sum_{p q, j  \neq k}  \hat{c}^\dagger_{k g p} \hat{c}^\dagger_{j e q} \hat{c}_{j g q} \hat{c}_{k e p}, 
	\ea
	where $j$ and $k$ are sites and $p$ and $q$ are nuclear spins. 
	The constants are given by $U_{gg} = 4 \pi \hbar a_{gg} \omega_\perp/L$, $U_{ee} =  4 \pi \hbar a_{ee} \omega_\perp/L$, $V = 4 \pi \hbar \frac{(a_{eg}^+ + a_{eg}^-)}{2}\omega_\perp/L$, and $V^{ex} = 4 \pi \hbar \frac{(a_{eg}^+ - a_{eg}^-)}{2}\omega_\perp/L$.
	Here $a_{eg}^+$ is the s-wave scattering length between atoms in a symmetric electronic $(g,e)$ configuration, $a_{eg}^-$ is the s-wave scattering length between atoms in an anti-symmetric electronic $(g,e)$ configuration and $a_{ee}$ is the s-wave scattering length between $e$ atoms.
	Note that $U_{gg}$ is written as $U$ in the main text for brevity.
	The evolution equations during the dark time are
	\ba
	\dot \rho_{\alpha \beta}^{m m'} &=& i \delta (\delta_{\alpha e} - \delta_{\beta e}) \rho_{\alpha\beta}^{m m'} \nonumber \\
	&&- i U_{gg} (n-1) \left[ \sum_r \delta_{\alpha g} (\rho_{\alpha \beta}^{m m'} \rho_{gg}^{rr} - \rho_{\alpha \beta}^{r m'} \rho_{gg}^{m r}) - \sum_r \delta_{\beta g} (\rho_{\alpha \beta}^{m m'} \rho_{gg}^{rr} - \delta_{\alpha \beta}^{m r} \rho_{gg}^{r m'})\right] \nonumber \\
	&&- i U_{ee} (n-1) \left[ \sum_r \delta_{\alpha e} (\rho_{\alpha \beta}^{m m'} \rho_{ee}^{rr} - \rho_{\alpha \beta}^{r m'} \rho_{ee}^{m r}) - \sum_r \delta_{\beta e} (\rho_{\alpha \beta}^{m m'} \rho_{ee}^{rr} - \delta_{\alpha \beta}^{m r} \rho_{ee}^{r m'})\right] \nonumber \\
	&& - i V (n-1) \left[ (\delta_{\alpha g} - \delta_{\beta g}) \sum_r \rho_{ee}^{rr} + (\delta_{\alpha e} - \delta_{\beta e}) \sum_r \rho_{gg}^{rr}\right] \rho_{\alpha \beta}^{m m'} \nonumber \\
	&& i V (n-1) \sum_r (\delta_{\alpha e} \rho^{r m'}_{g \beta} \rho^{m r}_{e g} - \delta_{\beta g} \rho^{m r}_{\alpha e} \rho^{r m'}_{e g} + \delta_{\alpha g} \rho^{r m'}_{e \beta} \rho^{m r}_{g e} - \delta_{\beta e} \rho^{m r}_{\alpha g} \rho^{r m'}_{g e}) \nonumber \\
	&& i V^{ex} (n-1) \sum_r (\delta_{\alpha g} \rho^{r m'}_{g \beta} \rho^{m r}_{e e} - \delta_{\beta g} \rho^{m r}_{\alpha g} \rho^{r m'}_{e e} + \delta_{\alpha e} \rho^{r m'}_{e \beta} \rho^{m r}_{g g} - \delta_{\beta e} \rho^{m r}_{\alpha e} \rho^{r m'}_{g g}) \nonumber \\
	&& - i V^{ex} (n-1) \sum_r (\delta_{\alpha e} \rho^{m m'}_{g \beta} \rho^{r r}_{e g} - \delta_{\beta g} \rho^{m m'}_{\alpha e} \rho^{r r}_{e g} + \delta_{\alpha g} \rho^{m m'}_{e \beta} \rho^{r r}_{g e} - \delta_{\beta e} \rho^{m m'}_{\alpha g} \rho^{r r}_{g e}).
	\ea
	Since there are no $m \neq m'$ components in the beginning of the dark time [see Eqs.\ (S1-S3)], we see that these components also stay zero during the dark time. The remaining evolution equations are
	\begin{eqnarray}
	\dot \rho_{gg}^{m m} &=& i V^{ex} (n-1) ( \rho^{m m}_{g e} \sum_r \rho^{r r}_{e g} -  \rho^{m m}_{e g} \sum_r \rho^{r r}_{g e}),\\
	\dot \rho_{ee}^{m m} &=& i V^{ex} (n-1)  (\rho^{m m}_{e g} \sum_r \rho^{r r}_{g e}- \rho^{m m}_{g e} \sum_r \rho^{r r}_{e g} ) \\
	&=& - \dot \rho_{gg}^{m m}, \\
	\dot \rho_{g e}^{m m} &=& - i \delta \rho_{ge}^{m m} - i U_{gg} (n-1)   \rho_{g e}^{m m} \sum_{r \neq m} \rho_{gg}^{rr} + i U_{ee} (n-1)   \rho_{g e}^{m m} \sum_{r \neq m} \rho_{ee}^{rr}  \nonumber \\
	&& - i V (n-1)  \rho_{g e}^{m m}  \sum_{r \neq m} (\rho_{ee}^{rr} -  \rho_{gg}^{rr}) - i V^{ex} (n-1) ( \rho^{m m}_{e e} -  \rho^{m m}_{g g}) \sum_{r \neq m} \rho^{r r}_{g e},
	\end{eqnarray}
	
	In terms of the matrix elements at the end of the dark time $\tau$, the measurement result (after the last pulse of area $-\beta$) is
	\ba
	\frac{\langle \hat n_e \rangle}{n} &=& \frac{1}{2} \left(1 + \sum_{m} (\rho_{ee}^{mm}(\tau) - \rho_{gg}^{mm}(\tau)) \cos \beta - i \sum_m (\rho_{eg}^{mm}(\tau) - \rho_{ge}^{mm}(\tau)) \sin \beta\right)  \\
	& \rightarrow &  \frac{1}{2} \left(1 -  \cos^2 \beta - i \sum_m (\rho_{eg}^{mm}(\tau) - \rho_{ge}^{mm}(\tau)) \sin \beta\right),
	\ea
	where the last line holds only for $\Gamma_{ee} = 0$, in which case the total number of $g$ atoms and total number of $e$ atoms are both conserved during the dark time (in $e$-$e$ losses, the total number of $e$ atoms is not conserved).

\section{EYD spectrum estimation}
\label{sec:EYDPr}

Here we calculate $\text{Pr}(\vec{\lambda}|n,\vec{p})$ exactly for finite $n$ for EYD measurement. 
This is required for Fig.~2(b) and Fig.~2(c) in the main text, along with the general calculation of $\langle \hat n_e(\tau,\vec{p}) \rangle / n $ via Eq.~(12), used to generate Fig.~2(a).

To carry out the analysis, first note that the measurement projectors  $\{ \Pi_{\vec{\lambda}} \}$ commute with the action of spin-rotation $\hat{V}^{\otimes n}$ applied to all spins. Therefore the measurement outcome is independent of the eigenstates of $\hat{\rho}$, and for the purpose of calculation we can take it to be $\hat{\rho} = \sum_{i=1}^N p_i |i \rangle \langle i |$. Thus the overall state of the system is
\begin{eqnarray}
\hat{\rho}^{\otimes n} &=& \left( \sum_{i=1}^N p_i |i \rangle \langle i |\right)^{\otimes n} \\
&=& \sum_{\mathclap{m_1,m_2,...,m_N | n }} p_1^{m_1} p_2^{m_2} ... p_N^{m_N} ~ \hat{M}_{m_1,m_2,...,m_N} ,
\end{eqnarray}
where the sum is over all non-negative integers $\{m_1,m_2,...,m_N \}$ such that $\sum_{i=1}^N m_i =n$, and $\hat{M}_{m_1,m_2,...,m_N}$ is the projector onto the subspace of states containing $m_i$ spin-state $i$'s (i.e. the state $|1\rangle^{\otimes m_1}|2\rangle^{\otimes m_2}...|N \rangle^{\otimes m_N}$, and all distinct permutations). Note that the subspace $\mathcal{H}_{m_1,m_2,...,m_N} \subset \mathcal{H}$ which $\hat{M}_{m_1,m_2,...,m_N}$ projects onto is preserved by the action of any permutation $\sigma \in S_n$, and therefore supports a representation of $S_n$. As such, $\mathcal{H}_{m_1,m_2,...,m_N}$ can be decomposed into irreps of $S_n$. For $c(\vec{\lambda} | \vec{m})$ copies of the $\vec{\lambda}$ irrep of $S_n$ in $\mathcal{H}_{m_1,m_2,...,m_N}$, the probability of obtaining measurement outcome $\vec{\lambda}$ is:
\begin{eqnarray}
\text{Pr}(\vec{\lambda}|n,\vec{p}) &=&  \text{Tr} \left( \Pi_{\vec{\lambda}} \hat{\rho}^{\otimes n} \right) =   \| \vec{\lambda}_{S_n} \| \sum_{\mathclap{m_1,m_2,...,m_N | n}} p_1^{m_1} p_2^{m_2} ... p_N^{m_N} ~ c(\vec{\lambda} | \vec{m}) ,
\label{eq:EYDPr}
\end{eqnarray}
where we remind the reader that $\Pi_{\vec{\lambda}}$ is the projector onto the subspace $\mathcal{H}_{\vec{\lambda}} \subset \mathcal{H}$, which carries the $\vec{\lambda}$-irrep of $S_n \times SU(N)$. Defining $l_i:= \lambda_i +N-i$, the dimension of the $\vec{\lambda}$ irrep of $S_n$ is
\begin{equation}
\label{DimensionIrrepSymmetric}
\| \vec{\lambda}_{S_n} \| = \frac{n!}{l_1 ! \cdots l_N!}\prod_{i<j}^N (l_i - l_j),
\end{equation}
which can be calculated directly for a particular instance $\vec{\lambda}$.

To obtain $c(\vec{\lambda} | \vec{m})$, first note that $c(\vec{\lambda} | \vec{m})$ cannot depend on the ordering of the integers in $\vec{m} =(m_1,m_2,...,m_N)$. Therefore it is sufficient to consider $c(\vec{\lambda} | \vec{\mu})$ for which $\mu_i \geq \mu_{i+1}$, therefore $\vec{\mu}$ specifies a valid Young diagram. Consider filling the $n$ boxes of the Young diagram $\vec{\lambda}$ with integers. We call the resulting filled Young diagram a {\it semi-standard Young tableau} if and only if the numbers are non-decreasing across rows from left to right, and strictly increasing down columns. Then $c(\vec{\lambda} | \vec{\mu})$ is the {\it Kostka number} $K_{\lambda \mu}$ \cite{gupta88}, which is given by the number of distinct semi-standard Young tableaux that can be constructed by filling Young diagram $\vec{\lambda}$ with $\mu_1$ $1$'s, $\mu_2$ $2$'s etc. This can be calculated numerically for particular instances of $\vec{\lambda}$ and $\vec{\mu}$. 

In the special case of $N=2$, taking $(\lambda_1,\lambda_2) = (\frac{n}{2}+S,\frac{n}{2}-S)$, the expression for $\text{Pr}(\vec{\lambda}|n,\vec{p})=\text{Pr}(S|n,\vec{p})$ takes a simple form. In this case, $ \| \vec{\lambda}_{S_n} \| = {n \choose \lambda_1} - {n \choose \lambda_1+1} $ and $c(\vec{\lambda} | \vec{\mu})$ is zero for $\mu_1 > \lambda_1$, and unity for $\mu_1 \leq \lambda_1$. Therefore,
\begin{eqnarray}
\text{Pr}(S|n,\vec{p}) &=& \left[ {n \choose \frac{n}{2}+S} - {n \choose \frac{n}{2}+S+1} \right]  \sum_{m=\frac{n}{2}-S}^{\frac{n}{2}+S} p^m(1-p)^{n-m}  .
\end{eqnarray}
This is used to generate Fig.~2(b) in the main text.

\section{Evaluating $\text{Tr}  \left(  \rho^{\otimes n} \mathcal B _w   \right)$}

In the main text, we require the evaluation of $\text{Tr}  \left(  \rho^{\otimes n} \mathcal B _w   \right)$ where $\mathcal{B}_w := e^{i \alpha \sum_{j =1}^{n-1-w}(1-s_{jn}) }$ in order to calculate $\langle \hat n_e(\tau,\vec{p}) \rangle / n $ in Eq.~(10). We provide two approaches to analyze $\text{Tr}  \left(  \rho^{\otimes n} \mathcal B _w   \right) $. In the main text we cover the first approach, which uses group representation theory, and in Sec.~\ref{sec:RepTheoryApproach} we provide a technical step required for the proof which was omitted from the main text. In Sec.~\ref{sec:NonRepTheoryApproach} we give an alternative analysis of $\text{Tr}  \left(  \rho^{\otimes n} \mathcal B _w   \right)$ which we use to prove that the deviation of Eq.~(2) in the main text from the exact result is $\tilde{\mathcal{O}}(1/\sqrt{n})$.

\subsection{Using group representation theory}
\label{sec:RepTheoryApproach}

In the main text in the paragraphs following Eq.~(12) we prove that
\begin{equation}
\label{Bresult2}
\tr_{\lambda}  \left[ \mathcal{B}_w \right]= \sum_\xi  m(\lambda,\xi)  \sum_{r=1}^N \!\frac{\| \xi^{-r} \| }{\| \lambda \|} v_r(\xi ),
\end{equation}
where $v_r(\xi)$ is the eigenvalue of $\mathcal{B}_w$ on the irrep $\xi^{-r}$. 
Here we prove the claim in the main text that $v_r(\xi) = e^{i \alpha ( n-w-1)} e^{ -i \alpha(\xi_r - r)}$.
In order to compute $v_r(\xi)$, it is necessary to understand the irrep $\xi^{-r}$ of $S_{l-1}$ inside the irrep $\xi$ of $S_l$, where $l := n-w$.
To this end, we construct a series of spaces of tabloids.
Recall that given a Young diagram $\xi = (\xi_1,...,\xi_N)$ with $\sum_r \xi_r =l$, a {\it Young tableau} $t$ is formed by inserting integers in the boxes of $\xi$. Here we consider those Young tableaux with each number from $1$ to $l$ appearing in precisely one box of $\xi$.
A {\it tabloid} $\{ t \} $ is an equivalence class of Young tableaux $t$, where two tableaux are equivalent 
if one is obtained from another by permuting within each row.
In other words, if $A_t$ is the {\it group of all row-preserving permutations} of $t$,
then $\{ t \} = \{ \alpha t : \alpha \in A_t \}$.
The symmetric group $S_l$ acts on the set of all tabloids by permuting numbers;
it can be verified that $\{ \pi t \} = \{ \pi \alpha t \}$ for any $\alpha \in A_t$ and $\pi \in S_{l}$,
and hence the notation $\pi \{t\}$ makes sense.
Let $B_t$ be the {\it group of all column-preserving permutations} of $t$,
and define
\[
e_t = \sum_{\beta \in B_t} \sgn(\beta) \beta \{ t \},
\]
which is called a {\it polytabloid}.
The action of $S_{l}$ on the span of all polytabloids is isomorphic to the irrep $\xi$.
A basis for this irrep can be chosen to be $\{ e_t : t \text{ is a standard Young tableau} \}$.
(A {\it standard tableau} is one in which numbers are increasing in each row and column.)

Define $V_i$ to be the span of $e_t$ where $t$ is a standard Young tableau with $n$ in one of the rows $1,\ldots,i$.
Certainly, $V_1 \subseteq V_2 \subseteq \cdots V_N = \xi$.
Observe that $V_i$ is a representation space of $S_{l-1}$ because the position of the number $l$ is fixed by $S_{l-1}$.
It is known that $V_i/V_{i-1}$ is isomorphic to $\xi^{-i}$ \cite{James1978}. 
Define $h:=  \sum_{j =1}^{l-1}s_{jl} $.
Note that $h$ preserves each $V_i$,
because $V_i$ and its orthogonal complement contain distinct irreps of $S_{l-1}$,
and the projection $\Pi_{V_i}$ onto $V_i$ from $\xi$ can be
written by some element of $\mathbb C S_{l-1}$, 
which implies that $h$ commutes with the projector $\Pi_{V_i}$.

The eigenvalue $v_r$ is determined by $h e_t = u_r e_t + w$, with $v_r=\exp(-i \alpha u_r)$, for some $e_t \in V_r \setminus V_{r-1}$ and $w \in V_{r-1}$.
We will read off the coefficient of $\{ t \}$, where `$l$' is placed in the row $i$ of a standard tableau $t$.
(If it is not possible for such $t$ to be standard, then $V_r/V_{r-1} = 0$.)
Since
\begin{equation}
h e_t = \sum_{ \tau \in h,~ \beta \in B_t }
\sgn(\sigma) \tau \beta \{ t \} ,
\end{equation}
we see that the coefficient of $\{t\}$ in $h e_t$ is
\begin{align}
u_r 
&= \sum_{ \tau \in h,~ \beta \in B_t ~:~ \tau \beta \{t\} = \{t\} } \sgn(\beta) = \sum_{ \tau \in h,~ \beta \in B_t ~:~ \tau \beta \in A_t } \sgn(\beta) .
\end{align}
In order to make a nonzero contribution to the sum, $\tau$ must be a member of $B_t \cdot A_t$.
If both $\alpha \in B_t$ and $\beta \in A_t$ are nontrivial, then $\beta \alpha$ cannot be a transposition.
Thus, $\tau= \beta \alpha$ must be a member of either $A_t$, in which case $\sgn(\beta=1)=1$, or $B_t$,
in which case $\sgn(\beta) = \sgn(\tau) = -1$. There are $\xi_r-1$ terms of $h$ that belong to $A_t$,
and $r-1$ terms of $h$ that belong to $B_t$. Therefore,
\begin{equation}
u_r = (\xi_r-1)(+1) + (r-1)(-1) = \xi_r - r.
\end{equation}
As  $\mathcal{B}_w = e^{i \alpha \sum_{j =1}^{n-1-w}(1-s_{jn}) } = e^{i \alpha ( n-w-1)} e^{-i \alpha h}  $, we see that $v_r(\xi) = e^{i \alpha ( n-w-1)} e^{ -i \alpha(\xi_r - r)}$ as required.

\subsection{Using elementary analysis}
\label{sec:NonRepTheoryApproach}

Our goal here is to show that the large-$n$ form of $\langle n_e(\tau, \vec{p}) \rangle/n$ is that of the mean field result Eq.~(2) in the main text, with a deviation which decreases as $\tilde{\mathcal{O}}(1/\sqrt{n})$.
First we calculate the expectation value of a permutation operator $ P(\sigma)$, defined as
\begin{equation}
P(\sigma) = \sum_{y_1 = 1}^d \cdots \sum_{y_n=1}^d \ket{y_{\sigma^{-1}(1)}, y_{\sigma^{-1}(2)}, \ldots, y_{\sigma^{-1}(m)}} \bra{y_1, y_2, \ldots, y_m } ,
\end{equation}
for permutation $\sigma$.
Let $\sigma = \sigma_1 \sigma_2 \cdots \sigma_m $ be the decomposition into disjoint cycles.
Some $\sigma_j$ may be 1-cycle.
By $|\sigma_j|$ we denote the length of a cycle.
For example, we have $|(1)|=1$, $|(56)|=2$ and $|(245)| = 3$.
The following equation is simple and useful,
\begin{align}
\tr( P(\sigma) \rho^{\otimes m} ) = \prod_j \tr(\rho^{|\sigma_j|}).
\label{eq:Ppi}
\end{align}
This is particularly simple to evaluate, since $\tr(\rho^l) = \sum_{r=1}^d p_r^l$, for $(p_1,p_2, \dots, p_d)$ the spectrum of $\rho$.
To prove this, it suffices to verify that 
(i) $P(\sigma) = P(\sigma_1) P(\sigma_2) \cdots P(\sigma_n)$  where distinct $P(\sigma_j)$ are supported on disjoint tensor factors, and
(ii) if $\sigma=\sigma_1$ is a cycle of length $m$, then $\tr(P(\sigma) \rho^{\otimes m}) = \tr(\rho^m)$. 
The truth of (i) is evident.
For (ii), we may assume $\sigma = (123\cdots m)$. Then,
\begin{align*}
\tr(P(\sigma) \rho^{\otimes m} ) 
&= \sum_{\{y_j\}} \bra{y_1, y_2, \ldots, y_{m} } \rho^{\otimes m} \ket{ y_m, y_1, \ldots, y_{m-1}}  \\
&= \sum_{\{y_j\}} \rho_{y_1 y_m} \rho_{y_2 y_1} \cdots \rho_{y_{m} y_{m-1}} = \tr( \rho^m ) .
\end{align*}

Next, we proceed to evaluate $\text{Tr}  \left(  \rho^{\otimes n} \mathcal B _w   \right)$ where $\mathcal{B}_w := e^{i \alpha \sum_{j =1}^{n-1-w}(1-s_{jn}) } = e^{i \alpha ( n-w-1)} e^{- i \alpha \sum_{j =1}^{n-1-w}s_{jn} }$ by expanding the exponential.
Let $m = n-w$ and $z = -i (\m-1) \alpha$.
Hereafter in this section, we denote by $\langle \cdot \rangle : = \text{Tr}(\rho^{\otimes m} ~ \cdot ~ ) $ the expectation value with respect to $\rho^{\otimes \m}$, 
\begin{align}
\langle \mathcal B _w \rangle &= e^{i \alpha ( \m-1)}  \langle e^{ (z/(\m-1))  \sum_{j=1}^{\m-1} s_{j,\m} }  \rangle
=
\frac{e^{i \alpha ( \m -1)} }{\m} \sum_{k = 1}^\m \langle e^{(z/(\m-1)) \sum_{j \neq k}^\m s_{j,k}} \rangle \\
&=
e^{i \alpha ( \m - 1)}  
\sum_{l=0}^\infty \frac{z^l}{l!}  \left\langle 
\underbrace{\left( \frac{1}{\m(\m-1)^l} \sum_{j_1,j_2,\ldots, j_l \neq k}^\m s_{j_1,k} s_{j_2,k} \cdots s_{j_l,k}\right) }_{X_l} \right\rangle . 
\end{align}
The operator $X_l$ contains precisely $\m(\m-1)^l$ terms in the sum.
Each summand is some permutation operator $\sigma \in S_{\m}$, 
and $\langle X_l \rangle$ can be interpreted as the average value $\langle \sigma \rangle$ 
upon a random choice of $\sigma$ among $\m(\m-1)^l$ possibilities.
(This probability distribution has nothing to do with $\Pr(w | n, \beta) $ above.)
From Eq.~(\ref{eq:Ppi}), we know that $\langle \sigma \rangle$ depends only on the lengths of cycles in the
disjoint cycle decomposition of $\sigma$.
If $j_1, \ldots, j_l$ are all distinct, then $\sigma = (j_1 k) \cdots (j_l k) = (k j_l j_{l-1} \cdots j_1)$ is a cycle of length $l+1$,
and $\langle \sigma \rangle = \tr \rho^{l+1} $.
If $\m$ is sufficiently large, then this is the most typical case.
Indeed, the probability that the $j_1, \ldots, j_l$ are all distinct (i.e. the probability that one obtains $\sigma$ of length $l+1$) is
\[
p(\m,l) = \frac{(l+1)! \binom{\m}{l+1}}{\m(\m-1)^l} = \frac{(\m-1)(\m-2) \cdots (\m-l)}{(\m-1)^l} \ge 1 - \frac{l^2}{\m-1}.
\]
This allows us to bound the ``error'' 
\[
\Delta_l := |\langle X_l \rangle - \tr \rho^{l+1}| \le (1-p(\m,l)) \cdot \max_{\sigma: |\sigma| \le l } | \langle \sigma \rangle - \tr(\rho^{l+1})| \le \frac{2l^2}{\m-1} \cc{,}
\]
where we used the trivial normalization $\tr( \rho^{l+1} ) \le 1$ and $\langle \sigma \rangle \le 1$.
Therefore,
\begin{align}
\langle \mathcal B _w \rangle &= e^{i \alpha ( \m-1)}  \sum_{l=0}^\infty \frac{z^l}{l!} \langle X_l \rangle 
= e^{i \alpha ( \m-1)}  \left[ \sum_{l=0}^\infty \frac{z^l}{l!} \tr( \rho^{l+1} ) + \frac{2}{\m-1} \mathcal O \left( \sum_{l=0}^\infty \frac{|z|^l l^2}{l!} \right)  \right]   \nonumber \\
&= e^{i \alpha ( \m-1)}  \left[ \sum_{r=1}^d p_r e^{z p_r } + \mathcal O \left( \frac{\exp(|z|)}{\m} \right) \right] . \label{eq:A-convergence}
\end{align}
This proves that in the limit when $\m = n-w$ is large, for fixed $\m \alpha $,
\ba
\mathfrak{Re}\, \{ e^{i \delta \tau}   \text{Tr}( \rho^{\otimes n} \mathcal{B}_w  ) \} &\rightarrow& \underbrace{\sum_r p_r \cos[ \alpha ( \m - 1 )(1- p_r )+ \delta \tau]}_{\mathcal{C}_w} + \mathcal O ( \m^{-1} ),
\ea
where we also have defined $\mathcal{C}_w$. Recall that from Eq.~(10) in the main text, we must sum over $w$ according to the binomial distribution $\Pr(w | n, \beta )$ in order to obtain $\langle \hat n_e(\tau,\vec{p}) \rangle / n$. 
We then see that 
\begin{align}
| \mathcal{C}_w - \mathcal{C}_{w'} | \le |(n-1)\alpha| \cdot \frac{|w-w'|}{n-1} \cc{,}
\end{align}
which is implied by the Taylor series (mean-value theorem) with respect to $w$. 

Using the tail bound for binomial distribution 
\[
\sum_{w: |w - \bar w| > (n-1)\epsilon } \Pr(w | n, \beta ) \le 2 e^{-2 (n-1) \epsilon^2} ,
\]
we arrive at the proof of the convergence of $\langle n_e \rangle / n$ for large $n$:
\begin{align}
&\left| \frac{\langle \hat{n}_e \rangle}{n} - \frac{\sin^2 \beta}{2}\left[ 1-  \sum_{r=1}^d p_r \cos \left(\cos^2 \! \frac{\beta}{2} \alpha ( n - 1 )(1- p_r )+ \delta \tau \right) \right]  \right| 
\label{eq:ne-conv}
\\
&\le
\frac12  \sum_w \Pr(w|n,\beta) | \mathfrak{Re}\, \{ e^{i \delta \tau}   \text{Tr}( \rho^{\otimes n} \mathcal{B}_w  ) \}  - \mathcal{C}_{\bar w -1}| \\
&\le
2 e^{-2(n-1) \epsilon^2} + \frac12 \max_{w : |w -\bar w| \le (n-1)\epsilon} | \mathfrak{Re}\, \{ e^{i \delta \tau}   \text{Tr}( \rho^{\otimes n} \mathcal{B}_w  ) \} - \mathcal{C}_w | + | \mathcal{C}_w - \mathcal{C}_{\bar w -1}| \\
&\le
2 e^{-2(n-1) \epsilon^2} + \frac{\mathcal O( \exp( n \alpha ) )}{n} + 2 (n \alpha) \epsilon \quad \quad \text{for any } \epsilon > 0\\
& \le
\mathcal O \left( \frac{\exp( n \alpha )}{\sqrt{ n / \log n}} \right) \quad \quad \text{setting }\epsilon^2 = \frac{\log n}{n}.
\end{align}

Therefore we have shown that Eq.~(2) in the main text differs from the exact result by $\tilde{\mathcal{O}}(1/\sqrt{n})$ in the limit $n \rightarrow \infty$ while holding $n \alpha$ constant. 
Recall that the tilde above the $\mathcal{O}$ means we neglect logarithmic factors.
There are a few comments on the technical aspects of the analysis above.
If $\beta$ is sufficiently small such that $n \sin^2 \frac{\beta}{2}$ is a constant irrespective of $n$, then $1/\sqrt n$ scaling is improved to be $1/n$. 
This is because the binomial distribution has smaller relative deviation when the probability is small.

\section{Numerical calculation of $\langle n_e(\tau, \vec{p}) \rangle/n$}

Here we collect the equations necessary to calculate $\langle n_e(\tau, \vec{p}) \rangle/n$ for the convenience of the reader. This is used in the main text to generate plots, for example Fig.~2(a). We also show how to evaluate  $\langle n_e(\tau, \vec{p}) \rangle/n$ more efficiently for large $n$ approximately by taking advantage of the fact that it is calculated in terms of narrow distributions.

By substituting Eq.~(11) into Eq.~(10) in the main text,
\begin{eqnarray}
\frac{\langle  \hat{n}_e(\tau,\vec{p}) \rangle}{n} ~  =  \frac{ \sin^2 \!\beta}{2} 
\left[ \! 1 \! -  \!  \sum_{w=0}^{n-1} \Pr(w | n,\beta)~  \mathfrak{Re}\, \{ e^{i \delta \tau}   \sum_\lambda \text{Pr}(\vec{\lambda}|n,\vec{p}) ~\text{Tr}_\lambda( \mathcal{B}_w) \}  \right] , 
\end{eqnarray}
where $\Pr(w|n,\beta) := \binom{n-1}{w} \cos^{2(n-w-1)} \frac{\beta}{2} \sin^{2w} \frac{\beta}{2}$. In Eq.~(\ref{eq:EYDPr}) in Sec.~\ref{sec:EYDPr} we showed that
\begin{eqnarray}
\text{Pr}(\vec{\lambda}|n,\vec{p}) &=& \| \vec{\lambda}_{S_n} \| \sum_{\mathclap{m_1,m_2,...,m_\N | n}} p_1^{m_1} p_2^{m_2} ... p_\N^{m_\N} ~ c(\vec{\lambda} | \vec{m}) ,
\end{eqnarray}
where $c(\vec{\lambda} | \vec{\mu})$ is the Kostka number, given by the number of distinct semi-standard Young tableaux that can be constructed by filling Young diagram $\vec{\lambda}$ with $\mu_1$ $1$'s, $\mu_2$ $2$'s etc, and [repeating Eq.~(\ref{DimensionIrrepSymmetric})]  the irrep dimension is
\begin{equation}
\| \vec{\lambda}_{S_n} \| = \frac{n!}{l_1 ! \cdots l_N!}\prod_{i<j}^N (l_i - l_j),~~~~\text{with}~l_i:= \lambda_i +N-i.
\end{equation}
The final step is to substitute for $\text{Tr}_\lambda( \mathcal{B}_w)$ as in Eq.~(11) in the main text
\begin{eqnarray}
\!\!\! \text{Tr}_\lambda( \mathcal{B}_w)  = e^{i\alpha(\!n\!-\!w\!-\!1\!)}
\sum_\xi \frac{m(\lambda,\xi) \|\xi \|}{ \| \lambda \|}  \sum_{r=1}^\N \!\frac{\| \xi^{-r} \| }{\| \xi \|}
e^{-i\alpha(\!\xi_r \! - r \!)} , 
\end{eqnarray}
where the sum is over all irreps $\xi$ of $S_{n-w}$ and $\xi^{-r}$ is the irrep of $S_{n-w-1}$ defined by removing a box from the $r$-th row of irrep $\xi$ of $S_{n-w}$.
The multiplicity $m(\lambda,\xi)$ is calculated iteratively from the branching rules which state that the restriction of an irrep $\lambda$ of $S_l$ to $S_{l-1}$ consists of distinct irreps $\lambda^{-r}$ of $S_{l-1}$ with multiplicity 1. 
Therefore, $m(\lambda,\xi)$ is the number of distinct paths from $\lambda$ to $\xi$, where each step in a path is a Young diagram, with one box removed from the previous step.

In the main text we show how to calculate $\langle \hat n_e \rangle/n$ exactly, here we describe how to drop terms to improve the efficiency of the calculation without sacrificing much accuracy.
We assume that $d$ is held fixed, and that $n$ becomes large here.
In the main text, we introduced three probability distributions
$\Pr (\lambda|n,p)$, $\Pr (w|n,\beta)$, and $\Pr (\xi|w,\lambda)$,
all of which turn out to be unimodal for large $n$.
The first one $\Pr(\lambda|n,p)$ is concentrated
at $\lambda \simeq n \vec p$ with the deviation of $\| \vec \lambda / n - \vec p \|$ being $\mathcal O(n^{-\frac12})$
by the result of EYD algorithm~\cite{SpectrumEstimationKeyl,christandl06}. 
By retaining only terms within a few standard deviations of $\vec p$ the number of $\vec{\lambda}$ that need to be summed over drops from $\sim \frac{1}{nd}\left( \frac{e^2n}{d^2} \right)^d \sim n^{d-1}$ \cite{Rob11} to approximately $\mathcal{O}(n^{(d-1)/2})$. 
The second distribution $\Pr (w | n,\beta)$ is the familiar binomial distribution. 
By including only terms within a few standard deviations of the mean, $\bar{w} = (n-1) \sin^2 \frac{\beta}{2}$, we reduce the number of $w$ which are summed from $\sim n$ to $\mathcal O(n^{-\frac12})$.The third distribution $\Pr(\xi |w, \lambda)$ is concentrated at $\xi \simeq \frac{n-w}{n} \lambda$
with the deviation $\| \frac{\vec \xi}{n-w} - \frac{\vec \lambda}{ n} \|$ being $\mathcal O(n^{-\frac12})$. 
There are $\mathcal{O}(n^{(d-1)/2})$ terms within a few standard deviations of the mean, as opposed to (what we expect to be) the full $\sim n^{d-1}$ terms.
Together therefore, the total number of terms after excluding those which contribute negligibly is reduced from $\sim n^{2d-1}$ to $\mathcal{O}(n^{(2d-1)/2})$.

\section{Effects of imperfections}

In this section we describe the effects of two main types of imperfections on the proposal, namely deviation from an exact square-well potential, and particle loss.
We will rely on numerics to analyze these cases as many of the symmetries which rendered our analysis tractable do not apply. 
For simplicity we consider there to be only two nuclear spin degrees of freedom, i.e., $d=2$.

First consider the case of a non-square well potential without loss. 
The Hamiltonian in Eq.~(1) of the main text is replaced by 
\begin{eqnarray}
\hat{H}_\textrm{D} =  \sum_{j < k} U_{jk} \hat \sigma_{gg}^{j} \hat \sigma_{gg}^{k}(1-\hat s_{jk}) - \delta \sum_k\hat{\sigma}_{ee}^k, \label{eq:HamiltonianUvaries}
\end{eqnarray}
where the strength of interaction has picked up mode dependence because the modes no longer are precise sinusoidal functions.
In Fig.~\ref{fig:ProbDist}(a) we plot $\langle \hat{n}_e(t) \rangle/n$ for each of the $n(n-1)/2$ constants $U_{jk}$ chosen uniformly from the interval $[U-d U/2, U+d U/2 ]$ for a variety of $\frac{d U}{U}$ ratios, where $\langle \hat{n}_e(t) \rangle$ is averaged over realizations. 
For the experimental parameters in the main text, i.e. with $L \sim 10 \mu$m, and using lasers with wavelength close to 600 nm, we can estimate two extremal values of $L_\pm \approx (10\pm 0.6) \mu$m. 
From the relation $U =(4 \pi a_{gg} \omega_\perp)/L$, we can thereby estimate $d U \approx (4 \pi a_{gg} \omega_\perp)/L_- -(4\pi a_{gg} \omega_\perp)/L_+ \approx 0.12 U$.
From Fig.~\ref{fig:ProbDist}(a) it is clear that the deviation in $n_e(t)$ due to $dU$ depends strongly on the time $t$. 
To estimate how much the typical $dU/U=0.12$ impacts the estimation of $p$, we therefore fix the time $t=1/U$ Fig.~\ref{fig:ProbDist}(b) shows the average $\langle \hat{n}_e(1/U) \rangle/n$, plus and minus its standard deviation (over realizations of $U_{jk}$ chosen uniformly from the interval $[U-d U/2, U+d U/2 ]$ for $dU/U=0.12$).
The largest deviations in the estimated $p$ occur near $p=1/2$, where an uncertainty of $\pm 0.05$ results from  $dU/U=0.12$.

Now consider the case of particle loss (but with $U_{jk}=U$ for all $j,k$ for simplicity). 
We write the evolution of the $n$-atom density matrix $\rho$ as,
\begin{eqnarray}
\dot \rho = -i [\hat{H}_D, \rho] - \frac{\Gamma}{2} \sum_{i<j} \left( \hat{c}_{ij}^\dagger \hat{c}_{ij} \rho + \rho \hat{c}_{ij}^\dagger \hat{c}_{ij} -2 \hat{c}_{ij} \rho \hat{c}_{ij}^\dagger \right), 
\label{eq:densitymatrixevolution}
\end{eqnarray}
where $\Gamma$ is the loss rate under lossy $e$-$e$ collisions \cite{zhang14}, and 
where $ c_{ij}$ is written in terms of atomic annihilation operators,
\begin{eqnarray}
\hat{c}_{ij} =\frac{1}{\sqrt{2}} \left( \hat{c}_{i e \downarrow} \hat{c}_{j e \uparrow} - \hat{c}_{i e \uparrow} \hat{c}_{j e \downarrow} \right).
\end{eqnarray}
For small $n$, one can calculate this evolution exactly; see Fig.~\ref{fig:ProbDist}(c).
In the (experimentally relevant) parameter regime of $\Gamma/U =0.5$, there is significant deviation compared with the loss-free case.

\begin{figure}[b]
	\includegraphics[width=0.95\textwidth]{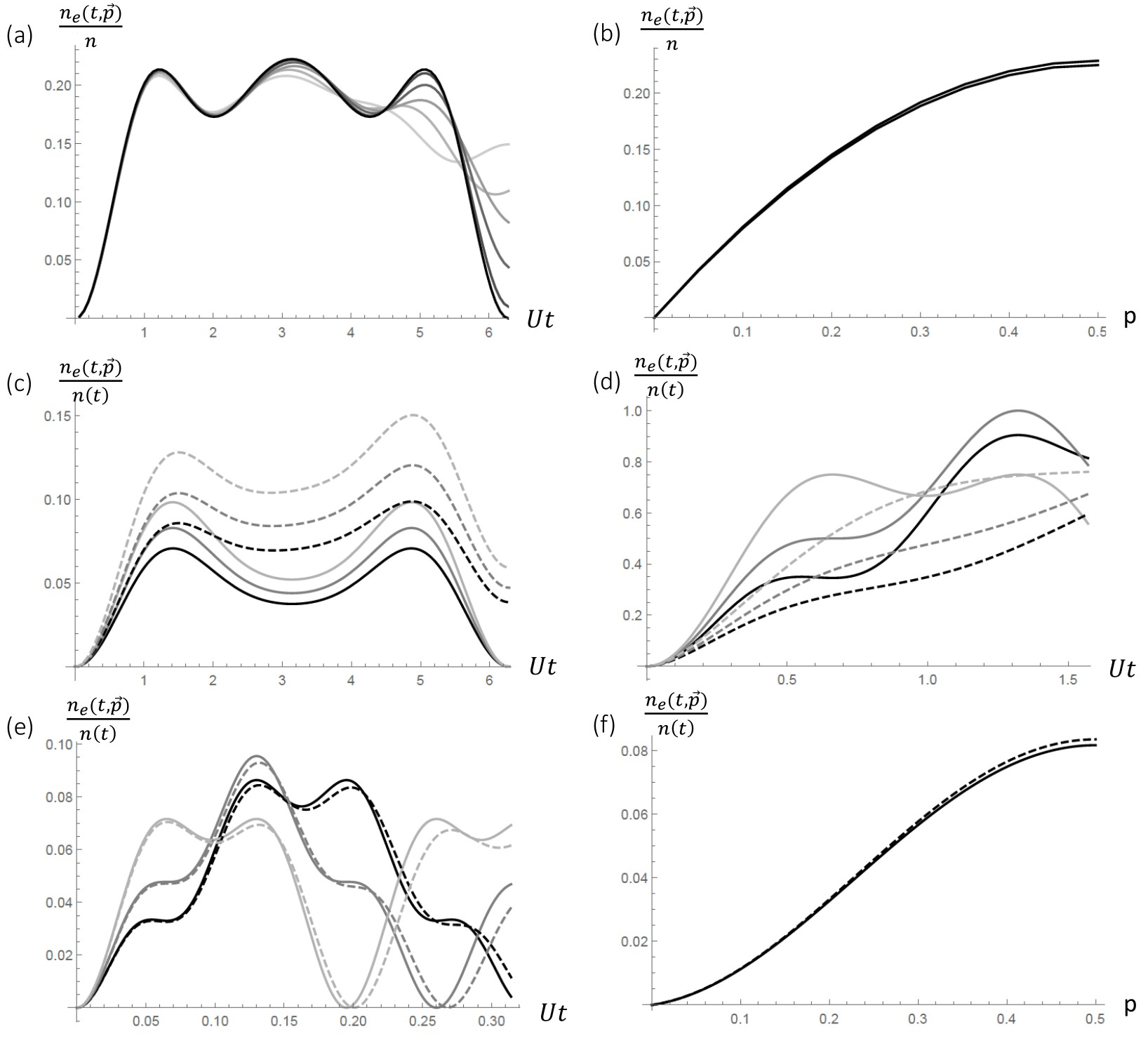}
	\caption{Plots of $n_e(t)/n(t) = \langle \hat n_e(t,\vec p) \rangle/\langle \hat n(t) \rangle$. 
		In all plots, $\delta =0$.
		(a) For $n=4$, and a variety of values of $d U/U=0,0.1,0.2,0.3,0.4,0.5$ (dark to light), using $\vec{p}=(2/3,1/3)$ and $\beta = \pi/4$.
		(b) To estimate the error that results from $dU/U=0.12$, we plot the mean $n_e(t)$, plus and minus the standard deviation over realizations at fixed time $t=1/U$ for $n=4$ as a function of $p$ in $\vec{p}=(1-p,p)$, with $\beta = \pi/4$ as in (a). 
		The largest uncertainties in estimating $p$ are expected to occur for $p$ near $1/2$ as the change in $n_e$ due to non-zero $dU$ is largest, and also the sensitivity of $n_e$ with respect to $p$ is least in that region.
		(c) For $n=4$, with $\Gamma/U=0$ (solid) and $\Gamma/U=0.5$ (dashed) using a variety of values of $\vec{p}=(4/5,1/5),(3/4,1/4),(2/3,1/3)$ (dark to light). Here, $\delta =0$ and $\beta = \pi/4$.
		(d) For $n=20$, with $\Gamma/U=0$ (solid) and $\Gamma/U=0.5$ (dashed) for a variety of values of $\vec{p}=(4/5,1/5),(3/4,1/4),(2/3,1/3)$ (dark to light). Here, $\delta =0$ and $\beta = \pi/4$.	
		The shape is altered significantly by $\Gamma$.
		(e) As in (d), but with $n=100$ and $\beta = \pi/20$.	
		The effect of $\Gamma$ is much less pronounced.
		(f) To estimate the error that results from $\Gamma/U=0.5$, we plot the mean $n_e(t)/n(t)$ for fixed time $t=0.05/U$ for $n=100$ as a function of $p$ in $\vec{p}=(1-p,p)$, and compare with the case for $\Gamma=0$ (dashed). 
		Here, $\beta = \pi/4$ as in (e)
		Those spectra with $p$ close to $1/2$ are most sensitive to loss.
	}
	\label{fig:ProbDist}
\end{figure}

To study loss for large $n$, we consider a mean-field approximation to Eq.~(\ref{eq:densitymatrixevolution}).
We remind the reader that the mean-field analysis is not valid for small $n$.
A part of this approximation is to assume the density matrix is separable,
\begin{eqnarray}
\rho = \otimes_{l=1}^n \left[\rho(l)_{in,in}\ket{in} \bra{in} + \rho(l)_{in,out} \ket{in} \bra{out} + \rho(l)_{out,in} \ket{out} \bra{in} + \rho(l)_{out,out} \ket{out} \bra{out}\right],
\end{eqnarray}
where we have introduced another degree of freedom $\{ \ket{in}, \ket{out} \}$ to track whether a particle is in the trap or has been lost, and $\rho(l)_{\alpha \beta}$ is a density matrix for a single atom $l$ with electronic and nuclear degrees of freedom.

Now consider taking the trace over all but the $j$th particle in the right hand side of Eq.~(\ref{eq:densitymatrixevolution}). The terms
$\bra{in} \text{Tr}_{n \setminus l}(-i [\hat{H}_D, \rho]) \ket{in}$, $\bra{in} \text{Tr}_{n \setminus l}(\hat{c}_{ij}^\dagger \hat{c}_{ij} \rho) \ket{in}$ and $\bra{in} \text{Tr}_{n \setminus l}(\rho \hat{c}_{ij}^\dagger \hat{c}_{ij}) \ket{in}$ have contributions only from density matrices $\rho(l)_{in,in}$ since $\hat{H}_D$ implicitly includes a projection onto atoms in the trap. 
Here, $\text{Tr}_{n \setminus l}(\cdot)$ implies tracing out the degrees of freedom on all atoms, except for atom $l$.
On the other hand, the term $\bra{in} \text{Tr}_{n \setminus l}(\hat{c}_{ij}\rho \hat{c}_{ij}^\dagger) \ket{in}$ must be zero since the recycling term outputs states in $\ket{out}$, which are cancelled by the projection $\bra{in} \cdot \ket{in}$.
Therefore Eq.~(\ref{eq:densitymatrixevolution}) becomes 
\begin{eqnarray}
\dot \rho(l)_{in,in} = \bra{in} \text{Tr}_{n \setminus l}(-i \hat{H}'\rho + i \rho \hat{H}'^\dagger) \ket{in}\!, \text{ where } \hat{H}' = - \delta \sum_k \hat{\sigma}_{ee}^k + \sum_{j<k} \left[ U \hat{\sigma}_{gg}^j \hat{\sigma}_{gg}^k (1-\hat{s}_{jk}) - \frac{i \Gamma}{4} \hat{\sigma}_{ee}^j \hat{\sigma}_{ee}^k (1-\hat{s}_{jk}) \right]\!\!. ~~~~
\label{eq:densitymatrixevolution2}
\end{eqnarray}
From here on, we drop the $in$ subscript on single-particle density operators.
Then,
\begin{eqnarray}
\dot \rho(l) &=& i \delta \left[ \hat{\sigma}_{ee}^l \rho(l) -\rho(l) \hat{\sigma}_{ee}^l \right]+\sum_{\mathclap{j<k (j \neq l, k \neq l)}} \frac{ -\Gamma}{2} A_e(j,k) \rho(l)+ \\
&+& \sum_{\mathclap{j=1 (j \neq l)}}^n -i U\left[B_g(l,k) - C_g(l,k)\right]   -\frac{\Gamma}{4}\left[ B_e(l,k)+ C_e(l,k) \right].\nonumber
\end{eqnarray}
where $U$ and $\Gamma$ are defined to be real, and where we define,
\begin{eqnarray}
A_\gamma(j,k) &=& \text{Tr}_{j k} \left[\hat{\sigma}_{\gamma \gamma}^j \hat{\sigma}_{ \gamma \gamma}^k(1-\hat{s}_{jk}) \rho(j) \otimes \rho(k) \right]=\text{Tr}_{j k} \left[  \rho(j) \otimes \rho(k) \hat{\sigma}_{\gamma \gamma}^j \hat{\sigma}_{ \gamma \gamma}^k(1-\hat{s}_{jk}) \right],\\
B_\gamma(l,k) &=& \text{Tr}_{k} \left[\hat{\sigma}_{\gamma \gamma}^l \hat{\sigma}_{ \gamma \gamma}^k(1-\hat{s}_{lk}) \rho(l) \otimes \rho(k) \right] = \text{Tr}_{k} \left[\hat{\sigma}_{\gamma \gamma}^k \hat{\sigma}_{ \gamma \gamma}^l(1-\hat{s}_{kl}) \rho(k) \otimes \rho(l) \right],\\
C_\gamma(l,k) &=& \text{Tr}_{k} \left[ \rho(l) \otimes \rho(k) \hat{\sigma}_{\gamma \gamma}^l \hat{\sigma}_{ \gamma \gamma}^k(1-\hat{s}_{lk}) \right]=\text{Tr}_{k} \left[ \rho(k) \otimes \rho(l) \hat{\sigma}_{\gamma \gamma}^k \hat{\sigma}_{ \gamma \gamma}^l(1-\hat{s}_{lk}) \right].
\end{eqnarray}
Each of these can be calculated explicitly,
\begin{eqnarray}
A_\gamma(j,k) &=& \sum_{mn} \rho_{\gamma \gamma}^{mm}(j) \rho_{\gamma \gamma}^{nn}(k) - \rho_{\gamma \gamma}^{nm}(j) \rho_{\gamma \gamma}^{mn}(k),\\
\left[ B_\gamma(l,k) \right]_{\eta \eta'}^{p p'} &=& \delta_{\eta \gamma} \sum_{n} \rho_{\gamma \eta'}^{p p'}(l) \rho_{\gamma \gamma}^{nn}(k) - \rho_{\gamma \eta'}^{np'}(l) \rho_{\gamma \gamma}^{pn}(k),\\
\left[ C_\gamma(l,k)  \right]_{\eta \eta'}^{p p'} &=& \delta_{\eta' \gamma}\sum_{n} \rho_{\eta \gamma}^{p p'}(l) \rho_{\gamma \gamma}^{nn}(k) - \rho_{\eta \gamma}^{pn}(l) \rho_{\gamma \gamma}^{np'}(k).
\end{eqnarray}
Finally, we use these to find the mean-field equations of motion in the case in which $\rho(l)$ is independent of $l$,
\begin{eqnarray}
\frac{d}{dt} \rho_{\eta \eta'}^{p p'} &=& i \delta \left( \delta_{\eta e}  \rho_{e \eta'}^{p p'} - \delta_{\eta' e} \rho_{\eta e}^{p p'} \right)
+
\frac{(n-1)(n-2)}{2} \left( \frac{ -\Gamma}{2} \right)  \rho_{\eta \eta'}^{p p'} \sum_{mn} \left( \rho_{e e}^{mm} \rho_{e e}^{nn} - \rho_{e e}^{nm} \rho_{e e}^{mn} \right) + \\
&-& i U(n-1)\left[ \delta_{\eta g} \sum_{n} \left( \rho_{g \eta'}^{p p'} \rho_{g g}^{nn} - \rho_{g \eta'}^{np'} \rho_{g g}^{pn} \right) - \delta_{\eta' g}\sum_{n} \left( \rho_{\eta g}^{p p'} \rho_{g g}^{nn} - \rho_{\eta g}^{pn} \rho_{g g}^{np'} \right) \right]+ \nonumber\\
&-&  \frac{\Gamma}{4}(n-1)\left[ \delta_{\eta e} \sum_{n} \left( \rho_{e \eta'}^{p p'} \rho_{e e}^{nn} - \rho_{e \eta'}^{np'} \rho_{e e}^{pn} \right) + \delta_{\eta' e}\sum_{n} \left( \rho_{\eta e}^{p p'} \rho_{e e}^{nn} - \rho_{\eta e}^{pn} \rho_{e e}^{np'} \right) \right].\nonumber
\end{eqnarray}
We use this to make the plots in Fig.~\ref{fig:ProbDist}(d), and observe that even for moderate $n =20$, non-zero $\Gamma/U=0.5$ alters the observed outcomes significantly.
Three possible approaches to overcome this problem are: (1) As described in the main text, reduce the radial trap strength for the excited atoms during the dark time. (2) Use a small $\beta$, which should help since the collisional effects arise at $\mathcal{O}(\beta^4)$, whereas the signal scales as $\sim \beta^2$. This has the downside of requiring more data to be taken to accommodate the reduced signal; Fig.~\ref{fig:ProbDist}(e). (3) Account for the modified evolution introduced by finite $\Gamma$ by including it in the model and using fits to the modified model to extract the spectrum. 
To estimate the uncertainty introduced in the estimation of $p$ by loss $\Gamma/U=0.5$ in the case in which a small tipping angle $\beta= \pi/20$ is used, we plot $\langle \hat{n}_e(t) \rangle/\langle \hat{n}(t) \rangle$ as a function of $p$ for fixed time $t=0.05/U$, and compare it with what would be expected if there was no loss in Fig.~\ref{fig:ProbDist}(f).
The largest deviations in the estimated $p$ occur near $p=1/2$, where a systematic shift of $- 0.05$ results from  $\Gamma/U=0.5$, however one could account for the corrections introduced by the known non-zero $\Gamma$.

\end{widetext}

\bibliography{SpectrumEstimationBibliography.bib}

\end{document}